\documentclass[aps]{revtex4}
\usepackage{amsfonts}
\begin{document}
\title{Acoustic Black Holes from Abelian Higgs Model with Lorentz Symmetry Breaking}
\author{M. A. Anacleto, F. A. Brito and E. Passos}
\email{anacleto, fabrito, passos@df.ufcg.edu.br}
\affiliation{Departamento de F\'{\i}sica, Universidade Federal de Campina Grande, 58109-970, Campina Grande, Para\'{\i}ba, Brazil}
\begin{abstract}
In this paper we derive acoustic black hole metrics in the (3+1) and (2+1)-dimensional Abelian Higgs model 
with Lorentz symmetry breaking. In this set up the sound waves lose the Lorentz boost invariance and suffer a `birefringence' effect.
We have found acoustic black holes and respective Hawking temperatures depending on the Lorentz violating parameter. Furthermore, we obtain an acoustic Kerr-like black hole metric with the Lorentz violating term affecting its rate of loss of mass. We also have shown that for suitable values of
the Lorentz violating parameter a wider spectrum of particle wave function can be scattered with increased amplitude by the acoustic black hole.

\end{abstract}
\maketitle
\pretolerance10000
\section{Introduction}

Acoustic black holes possess many of the fundamental properties of the black holes of general relativity and have been extensively studied in the literature \cite{Volovik, MV, Ok, Visser, Xian, Unruh}. The connection between black hole physics and the theory of supersonic acoustic flow was established in 1981 by Unruh \cite{Unruh} and has been developed to investigate the Hawking radiation 
and other phenomena for understanding quantum gravity. Hawking radiation is an important quantum effect of black hole physics. In 1974, Hawking combining Einstein's General Relativity and
Quantum Mechanics announced that classically a black hole does not radiate, but when we consider quantum effects emits thermal radiation at a temperature proportional to the horizon surface gravity.

Since the Hawking radiation showed by Unruh \cite{Unruh} is a purely
kinematic effect of quantum field theory, we can study the Hawking radiation process in completely
different physical systems. For example, acoustic horizons are regions where a moving fluid exceeds the local sound speed through 
a spherical surface and possesses many of the properties associated with
the event horizons of general relativity. In particular, the acoustic Hawking radiation when quantized appears as a flux of thermal phonons
emitted from the horizon at a temperature proportional to its surface gravity. Many fluid systems
have been investigated on a variety of analog models of acoustic black holes, including gravity wave~\cite{RS}, water
\cite{Mathis}, slow light \cite{UL}, optical fiber \cite{Philbin} and  electromagnetic waveguide \cite{RSch}. The models of superfluid helium II \cite{Novello}, atomic Bose-Einstein condensates \cite{Garay,OL} and one-dimensional Fermi degenerate
noninteracting gas \cite{SG} have been proposed to create an acoustic black hole geometry in
the laboratory.

The purpose of this paper is considering the idea of the Lorentz symmetry breaking theories suggested in the seminal paper in Superstring Theory \cite{Kost} and further developed in Quantum Field Theory and General Relativity \cite{Colladay:1998fq, Kostelecky:2003fs, CFJ, JAK-PI} to investigate the relativistic version of acoustic black holes from the Abelian Higgs model \cite{Xian} with Lorentz symmetry breaking. 

The main novel results we found are the following.
Differently of the most cases studied, we consider a relativistic fluid plus a term that violates the Lorentz symmetry. 
We derive acoustic black hole metrics in the (3+1) and (2+1)-dimensional Abelian Higgs model 
with Lorentz symmetry breaking. The effects of this
set up is such that the fluctuations of the fluids are also affected. The sound waves inherit the broken Lorentz symmetry of the fluid, lose the Lorentz boost invariance and develops a birefringence phenomenon. As consequence the Hawking temperature is directly affected by the Lorentz-violating 
term. Analogously to Lorentz-violating gravitational black holes \cite{syb,adam}, the effective Hawking temperature of the acoustic black holes 
now is {\it not} universal for all species of particles. It
depends on the maximal attainable velocity of this species.
 Furthermore, the acoustic black hole metric can be identified with 
an acoustic Kerr-like black hole. The Lorentz violating term affects the rate of loss of mass of the black hole. We also have shown that for suitable values of
the Lorentz violating parameter a wider spectrum of particle wave function can be scattered with increased amplitude by the acoustic black hole. This
increases the superressonance phenomenon previously studied in \cite{Basak:2002aw}.

The paper is organized as follows. In Sec.~\ref{II} we present the extended Abelian Higgs model with the Lorentz-violating term to find the acoustic 
black holes metrics. In Sec.~\ref{III} we study the dispersion relation. We find that the sound speed propagates with speeds lower or higher than its
usually maximal attainable speed in the fluid. In Sec.~\ref{IV} we study the acoustic black holes in their canonical form by rewriting the metric in spherical coordinate in a Schwarzschild and Kerr-like form. In Sec.~\ref{V} we briefly discuss on the (2+1) dimensional Abelian Higgs model. In Sec.~\ref{conclu} we make our final conclusions.

\section{The Lorentz-Violating Model}
\label{II}
In this section we make an extension of the Abelian Higgs model by modifying its scalar sector with a scalar Lorentz-violating term \cite{Bazeia:2005tb}.
Thus, the Lagrangian of the Lorentz-violating Abelian Higgs model is
\begin{eqnarray}
\label{acao}
{\cal L}&=&-\frac{1}{4}F_{\mu\nu}F^{\mu\nu} +|D_{\mu}\phi|^2+ m^2|\phi|^2-b|\phi|^4+ k^{\mu\nu}D_{\mu}\phi^{\ast}D_{\nu}\phi,
\end{eqnarray}
where $F_{\mu\nu}=\partial_{\mu}A_{\nu}-\partial_{\nu}A_{\mu}$, $D_{\mu}\phi=\partial_{\mu}\phi - ieA_{\mu}\phi$ and $k^{\mu\nu}$  is a constant tensor implementing the Lorentz symmetry breaking. The tensor coefficient is a $4\times4$ matrix, given by
\begin{equation}
k_{\mu\nu}=\left[\begin{array}{clcl}
\beta &\vdots & \alpha\\
\cdots &\cdot &\cdots\\
\alpha &\vdots & \beta
\end{array}\right], \quad(\mu,\nu=0,1,2,3),
\end{equation}
with $\alpha$ and $\beta$ being real parameters. 

\subsection{The case $\beta\neq 0$ and $\alpha=0$}
\noindent
Now let us use the decomposition $\phi=\sqrt{\rho(x,t)}\exp{(iS(x,t))}$ in the original Lagrangian
\begin{eqnarray}
{\cal L}&=&-\frac{1}{4}F_{\mu\nu}F^{\mu\nu}+\rho\partial_{\mu}S\partial^{\mu}S
-2e\rho A_{\mu}\partial^{\mu}S + e^2\rho A_{\mu}A^{\mu} + m^2\rho-b\rho^2
\nonumber\\
&+&k^{\mu\nu}\rho(\partial_{\mu}S\partial_{\nu}S-2eA_{\mu}\partial_{\nu}S+ e^2 A_{\mu}A_{\nu})
+\frac{\rho}{\sqrt{\rho}}(\partial_{\mu}\partial^{\mu}+k^{\mu\nu}\partial_{\mu}\partial_{\nu})\sqrt{\rho}.
\end{eqnarray}
The corresponding equations of motion are:
\begin{eqnarray}
\label{cont}
&&-\partial_{t}\left[\tilde{\beta}_{+}\rho(\dot{S}-eA_{t})\right]
+\partial_{i}\left[\tilde{\beta}_{-}\rho(\partial^{i}S-eA^{i})\right]=0, \quad\quad i=1,2,3
\\
\label{fluid}
&&\frac{(\tilde{\beta}_{+}\partial^2_{t}-\tilde{\beta}_{-}\partial^2_{i})\sqrt{\rho}}{\sqrt{\rho}}
+\tilde{\beta}_{+}\left(\dot{S}-eA_{t}\right)^2
-\tilde{\beta}_{-}\left(\partial_{i}S-eA_{i}\right)^2
+m^2-2b\rho=0,
\end{eqnarray}
where $\tilde{\beta}_{\pm}\equiv 1\pm\beta$ and for simplicity we first take $\alpha=0$. 

For the field $A_{\mu}$, we obtain the modified 
Maxwell's equations
\begin{eqnarray}
\partial_{\mu}F^{\mu\nu}=2e\rho\left[\partial^{\nu}S-eA^{\nu}+k^{\nu\mu}(\partial_{\mu}S-eA_{\mu})\right],
\end{eqnarray}
that is, there exist changes in the Gauss and Amp\`ere laws
\begin{eqnarray}
\nabla\cdot\vec{E}=2e\rho(1+\beta)w,
\\
\nabla\times\vec{B}-\partial_{t}\vec{E}=2e\rho(1-\beta)\vec{v},
\end{eqnarray}
where we have defined $w=-\dot{S}+eA_{t}$ and $\vec{v}=\nabla S+e\vec{A}$.
Similar changes have also appeared in \cite{JAK-PI,CFJ}.

The Eq.~(\ref{cont}) is the continuity equation and Eq.~(\ref{fluid}) is an equation describing a hydrodynamical fluid with a term 
$\frac{(\tilde{\beta}_{+}\partial^2_{t}-\tilde{\beta}_{-}\partial^2_{i})\sqrt{\rho}}{\sqrt{\rho}}$ called quantum potential (quantum correction term)
\cite{MV}. Now we consider perturbations around the background $(\rho_{0}, S_{0})$, with $\rho=\rho_{0}+ \rho_{1}$ and $S=S_{0}+S_{1}$, so we can rewrite (\ref{cont}) and (\ref{fluid}) as:
\begin{equation}
-\partial_{t}\left[\tilde{\beta}_{+}\rho_{0}\dot{S}_{1}+\tilde{\beta}_{+}\rho_{1}\left(\dot{S}_{0}-eA_{t}\right)\right]
+\partial_{i}\left[\tilde{\beta}_{-}\rho_{0}\partial^{i}S_{1}
+\tilde{\beta}_{-}\rho_{1}\left(\partial^{i}S_{0}-eA^{i}\right)\right]=0,
\end{equation}
\begin{equation}
2\tilde{\beta}_{+}\left(\dot{S}_{0}-eA_{t}\right)\dot{S}_{1}
-2\tilde{\beta}_{-}\left(\partial_{i}S_{0}-eA_{i}\right)\partial_{i}S_{1} - b\rho_{1} 
+(\tilde{\beta}_{+}D_{t2}+\tilde{\beta}_{-}D_{i2})\rho_{1}=0,
\end{equation}
where
\begin{eqnarray}
&&D_{{t}2}\rho_{1}=-\frac{1}{2}\rho^{-\frac{3}{2}}_{0}(\partial^2_{{t}}\sqrt{\rho_{0}})\rho_{1}
+\frac{1}{2}\rho^{-\frac{3}{2}}_{0}\partial^2_{{t}}(\rho_{0}^{-\frac{1}{2}}\rho_{1}),
\nonumber\\
&&D_{{i}2}\rho_{1}=-\frac{1}{2}\rho^{-\frac{3}{2}}_{0}(\partial_{i}\partial^{i}\sqrt{\rho_{0}})\rho_{1}
+\frac{1}{2}\rho^{-\frac{3}{2}}_{0}\partial_{i}\partial^{i}(\rho_{0}^{-\frac{1}{2}}\rho_{1}).
\end{eqnarray}
Thus, the wave equation for the perturbations $S_1$ around the background $S_0$ becomes
\begin{eqnarray}
\label{eqwave}
&&\partial_{t}\left[\left(-\frac{b\rho_{0}}{2\tilde{\beta_{-}}}
+\frac{{\cal D}_{2}\rho_{0}}{2\tilde{\beta_{-}}}-\frac{\tilde{\beta}_{+}}{\tilde{\beta}_{-}}w_{0}^2\right)\dot{S}_{1}
-w_{0}\vec{v}_{0}\cdot \nabla S_{1}\right]
\nonumber\\
&&+\nabla\cdot\left[-w_{0}\vec{v}_{0}\dot{S}_{1}+\left(\frac{b\rho_{0}}{2\tilde{\beta}_{+}}-\frac{{\cal D}_{2}\rho_{0}}{2\tilde{\beta_{+}}}\right)\nabla S_{1}
-\frac{\tilde{\beta}_{-}}{\tilde{\beta}_{+}}\vec{v}_{0}\cdot \nabla S_{1}\vec{v}_{0}\right]=0,
\end{eqnarray}
where we have defined, ${\cal D}_{2}=\tilde{\beta}_{+}D_{t2}+\tilde{\beta}_{-}D_{i2}$, $w_{0}=-\dot{S}_{0}+eA_{t}$ and $\vec{v}_{0}=\nabla S_{0}+e\vec{A}$ (the local velocity field). 

This equation can be seen as the Klein-Gordon equation in a curved space in (3+1) dimensional and can be written as~\cite{Unruh}
\begin{equation}
\label{eqkg}
\frac{1}{\sqrt{-g}}\partial_{\mu}\sqrt{-g}g^{\mu\nu}\partial_{\nu}S_{1}=0,
\end{equation}
where the idex $\mu=t,x_{i}$, $i=1,2,3$ and being the metric components given in the form
\begin{equation}
\sqrt{-g}g^{\mu\nu}\equiv\left[\begin{array}{clcl}
-\frac{b\rho_{0}}{2\tilde{\beta_{-}}}+\frac{{\cal D}_{2}\rho_{0}}{2\tilde{\beta_{-}}}
-\frac{\tilde{\beta}_{+}}{\tilde{\beta}_{-}}w^2_{0}&\vdots & \hskip1.5cm -w_{0}v^{j}_{0}\\
\cdots\cdots\cdots\cdots\cdots\cdots\cdots & \cdot & \quad\quad\cdots\cdots\cdots\cdots\cdots\cdots\\
-w_{0}v^{i}_{0} &\vdots& \hskip0,5cm \left(\frac{b\rho_{0}}{\tilde{\beta}_{+}}-\frac{{\cal D}_{2}\rho_{0}}{2\tilde{\beta_{+}}}\right)\delta^{ij}
-\frac{\tilde{\beta}_{-}}{\tilde{\beta}_{+}}v^{i}v^{j}
\end{array}\right].
\end{equation}

In terms of the inverse of $g^{\mu\nu}$ we have the metric of an acoustic black hole
\begin{equation}
g_{\mu\nu}\equiv\frac{b\rho_{0}\tilde{\beta}^{1/2}_{-}}{2c_{s}\sqrt{{\cal Q}(c^2_{s},v^2)}}
\left[\begin{array}{clcl}
-\left(\frac{c_{s}^2}{\tilde{\beta}_{+}}-\frac{{\cal D}_{2}c^2_{s}}{b\tilde{\beta_{+}}}-\frac{\tilde{\beta}_{-}}{\tilde{\beta}_{+}}v^2\right)& \vdots &\quad\quad\quad\quad-v^{j}\\
\cdots\cdots\cdots\cdots\cdots &\cdot & \quad\quad\cdots\cdots\cdots\cdots\cdots\cdots\\
-v^{i} & \vdots & f(c^2_{s},v^2)\delta^{ij}+\frac{\tilde{\beta}_{-}}{\tilde{\beta}_{+}}v^{i}v^{j}
\end{array}\right],
\end{equation}
where we have defined the local sound speed in the fluid as $c_{s}^2=\frac{b\rho_{0}}{2w^{2}_{0}}$; $v^{i}=\frac{v_{0}^i}{w_{0}}$, $v^2=\sum_{i}v^{i}v^{i}$ $(i=1,2,3$),
$f(c^2_{s},v^2)=\frac{\tilde{\beta}_{+}}{\tilde{\beta}_{-}}+\frac{c_{s}^2}{\tilde{\beta}_{-}}-\frac{{\cal D}_{2}}{b\tilde{\beta}_{-}}c^2_{s} 
-\frac{\tilde{\beta}_{-}}{\tilde{\beta}_{+}}v^2$ and 
${\cal Q}(c^2_{s},v^2)=1+\frac{c_{s}^2}{\tilde{\beta}_{+}}-\frac{\tilde{\beta}_{-}}{\tilde{\beta}_{+}}v^2
-\frac{{\cal D}_{2}}{b}-\frac{2c^2_{s}{\cal D}_{2}}{b\tilde{\beta}_{+}}+\frac{c^2_{s}{\cal D}^2_{2}}{b\tilde{\beta}_{+}}
+\frac{\tilde{\beta}_{-}{\cal D}_{2}v^2}{b\tilde{\beta}_{+}}$. One can make this metric simpler, since the parameter ${\cal D}_{2}$ is small enough to allowing us the dropping of the corresponding term out 
\begin{equation}
\label{invs_g}
g_{\mu\nu}\equiv\frac{b\rho_{0}\tilde{\beta}_{-}^{1/2}}{2c_{s}\sqrt{1+\frac{c_{s}^2}{\tilde{\beta}_{+}}
-\frac{\tilde{\beta}_{-}}{\tilde{\beta}_{+}}v^2}}
\left[\begin{array}{clcl}
-\left(\frac{c_{s}^2}{\tilde{\beta}_{+}}-\frac{\tilde{\beta}_{-}}{\tilde{\beta}_{+}}v^2\right)& \vdots &\quad\quad\quad\quad-v^{j}\\
\cdots\cdots\cdots\cdots\cdots &\cdot & \quad\quad\cdots\cdots\cdots\cdots\cdots\cdots\\
-v^{i} & \vdots & \left(\frac{\tilde{\beta}_{+}}{\tilde{\beta}_{-}}+\frac{c_{s}^2}{\tilde{\beta}_{-}}
-\frac{\tilde{\beta}_{-}}{\tilde{\beta}_{+}}v^2\right)\delta^{ij}+\frac{\tilde{\beta}_{-}}{\tilde{\beta}_{+}}v^{i}v^{j}
\end{array}\right].
\end{equation}
The metric 
depends simply on the density $\rho_0$, the local sound speed in the fluid $c_s$, the velocity of flow $\vec{v}$ and the parameter $\tilde{\beta}_{\pm}=(1\pm\beta)$. The latter is responsible for breaking the Lorentz symmetry. 
Notice that the sound speed $c_{s}$ is a function of the electromagnetic field $A_{t}$.
The acoustic line element can be written as
\begin{eqnarray}
ds^2&=&\frac{b\rho_{0}\tilde{\beta}_{-}^{1/2}}{2c_{s}\sqrt{\cal{Q}}}
\left[-\left(\frac{c_{s}^2}{\tilde{\beta}_{+}}-\frac{\tilde{\beta}_{-}}{\tilde{\beta}_{+}}v^2\right)dt^2-2\vec{v}\cdot d\vec{x}dt
+\frac{\tilde{\beta}_{-}}{\tilde{\beta}_{+}}(\vec{v}\cdot d\vec{x})^2
+f_{\beta} d\vec{x}^2\right],
\end{eqnarray}
where ${\cal Q}=1+\frac{c_{s}^2}{\tilde{\beta}_{+}}-\frac{\tilde{\beta}_{-}}{\tilde{\beta}_{+}}v^2$
and $f_{\beta}=\frac{\tilde{\beta}_{+}}{\tilde{\beta}_{-}}+\frac{c_{s}^2}{\tilde{\beta}_{-}}-\frac{\tilde{\beta}_{-}}{\tilde{\beta}_{+}}v^2$. Now changing the time coordinate as $d\tau=dt + \frac{\tilde{\beta}_{+}\vec{v}\cdot d\vec{x}}{c^2_{s}-\tilde{\beta}_{-}v^2}$ we find the acoustic metric in the stationary form
\begin{eqnarray}
ds^2&=&\frac{b\rho_{0}\tilde{\beta}_{-}^{1/2}}{2c_{s}\sqrt{{\cal Q}}}
\left[-\left(\frac{c_{s}^2}{\tilde{\beta}_{+}}-\frac{\tilde{\beta}_{-}}{\tilde{\beta}_{+}}v^2\right)d\tau^2+
{\cal F}\left(\frac{\tilde{\beta}_{-}v^{i}v^{j}}{c^2_{s}-\tilde{\beta}_{-}v^2}
+\frac{f_{\beta}}{{\cal F}}\delta^{ij}\right)dx^{i}dx^{j}\!\right]\!\!.
\end{eqnarray}
where ${\cal F}=\left(\frac{\tilde{\beta}_{+}}{\tilde{\beta}_{-}}+\frac{c_{s}^2}{\tilde{\beta}_{+}}
-\frac{\tilde{\beta}_{-}}{\tilde{\beta}_{+}}v^2\right)$.
This is the black hole metric for high $c_s$ and $\vec{v}$ speeds. It would be relativistic were the Lorentz violating parameter $\tilde{\beta}=1$ as in Ref.~\cite{Xian}.
In this case the corresponding Hawking temperature at the event horizon is given by 
\begin{equation}
T_{H}=\frac{1}{2\pi\tilde{\beta}_{+}}\frac{\partial(c_{s}-\tilde{\beta}_{-}v^i)}{\partial x_{i}}\Big|_{horizon}.
\end{equation}
Notice that the parameter $\tilde{\beta}$ plays a non-trivial role on the Hawking temperature. 

\subsection{The case $\beta=0$ and $\alpha\neq 0$}
\noindent
In this case, for the field $A_{\mu}$, the changes in the Gauss and Amp\`ere laws are now given by
\begin{eqnarray}
\nabla\cdot\vec{E}=2e\rho[w_{0}+\alpha w_{0}-\alpha v_{0}],
\\
(\nabla\times\vec{B})^{i}-\partial_{t}E^{i}=2e\rho[v^{i}_{0}+k^{i0}w_{0}-k^{ij}v_{0j}],
\end{eqnarray}
just as in the previous case and also discussed in \cite{JAK-PI,CFJ}.

Following the same steps described for the previous case, we consider perturbations around the background $(\rho_{0}, S_{0})$, with $\rho=\rho_{0}+ \rho_{1}$ and $S=S_{0}+S_{1}$, so, the wave equation for the perturbations $S_1$ around the background $S_0$ becomes
\begin{eqnarray}
\label{eqkga}
-\partial_{t}\left[\frac{b\rho_{0}}{2c^{2}_{s}}(a^{tt}\dot{S}_{1}
+a^{ti}\partial_{i}S)\right]+\partial_{i}\left[\frac{b\rho_{0}}{2c^{2}_{s}}(a^{it}\dot{S}_{1}+a^{ij} \partial_{j}S_{1})\right]=0,
\end{eqnarray} 
where
\begin{eqnarray}
a^{tt}&=&\left[c^2_{s}+(1-\alpha v)^2\right],
\\
a^{ti}&=&(1-\alpha v)v^{i},
\\
a^{it}&=&-(1-\alpha v)v^{i},
\\
a^{ij}&=&[c^2_{s}(1+\alpha)+\alpha^{2}(1-v)^2]\delta^{ij}-v^{i}v^{j}.
\end{eqnarray}
This relation can be written as a Klein-Gordon equation type (\ref{eqkg}) and we have the metric of an acoustic black hole given in the form
\begin{equation}
\label{met}
g_{\mu\nu}\equiv\frac{b\rho_{0}}{2c_{s}\sqrt{f}}
\left[\begin{array}{clcl}
g_{tt} &\vdots &g_{tj}\\
\cdots&\cdot&\cdots\\
g_{it} &\vdots &g_{ij}
\end{array}\right],
\end{equation}
where
\begin{eqnarray}
\label{gtt}
g_{tt}&=&-[(1+\alpha)c^2_{s}-v^2+\alpha^2(1-v)^2],
\\
g_{tj}&=&-(1-\alpha v)v^{j},
\\
g_{it}&=&-(1-\alpha v)v^{i},
\\
g_{ij}&=&\left[(1-\alpha v)^2+c^2_{s}-v^2\right]\delta^{ij} +v^{i}v^{j},
\label{gij}
\\
f&=&(1+\alpha)[(1-\alpha v)^2+c^2_{s}]-v^2+\alpha^{2}(1-v)^2\left[1+(1-\alpha v)^2c^{-2}_{s}\right].
\end{eqnarray}
Thus, the acoustic line element can be written as
\begin{eqnarray}
ds^2&=&\frac{b\rho_{0}}{2c_{s}\sqrt{f}}
\left[g_{tt}dt^2-2(1-\alpha v)(\vec{v}\cdot d\vec{x})dt+(\vec{v}\cdot d\vec{x})^2
+\left[(1-\alpha v)^2+c^2_{s}-v^2\right] d\vec{x}^2\right].
\end{eqnarray}
Now changing the time coordinate as 
$d\tau=dt + \frac{(1-\alpha v)(\vec{v}\cdot d\vec{x})}{[(1+\alpha)c^2_{s}-v^2+\alpha^2(1-v)^2]}$,
we find the acoustic metric in the stationary form
\begin{eqnarray}
ds^2=\frac{b\rho_{0}}{2c_{s}\sqrt{f}}
\left[g_{tt}d\tau^2+[(1-\alpha v)^2-g_{tt}]\left(\frac{-v^{i}v^{j}}{g_{tt}}
+\frac{\left[(1-\alpha v)^2+c^2_{s}-v^2\right]\delta^{ij}}
{[(1-\alpha v)^2-g_{tt}]}\right)dx^{i}dx^{j}\right]\!.
\end{eqnarray}
For $\alpha=0$, the result in \cite{Xian} is recovered. 

\section{The Dispersion Relation}
\label{III}
The sound waves are usually governed by an effective Lorentzian spacetime geometry. 
In order to study the effect of the Lorentz violating term in such structure we should investigate the dispersion relation. 
So let us now discuss the dispersion relation for the cases discussed above.
\subsection{The case $\beta\neq0$ and $\alpha= 0$}
\noindent
We now derive the dispersion relation from the equation (\ref{eqwave}). Since the field $S_{1}$ is real we use the notation
\begin{equation} 
S_{1}\sim\mbox{Re}[e^{(-i\omega t + i\vec{k}\cdot\vec{x})}],\quad
\omega=\frac{\partial S_{1}}{\partial t}, \quad k_{i}=\nabla_{i} S_{1}.
\end{equation}
In this case, the Klein-Gordon equation (\ref{eqwave}) in terms of momenta and frequency, becomes
\begin{equation}
\tilde{\beta}_{+}(c^2_{s}+\tilde{\beta}_{+})\omega^2+2\tilde{\beta}_{-}\tilde{\beta}_{+}(\vec{v}\cdot\vec{k})\omega
-\tilde{\beta}_{-}(c^2_{s}-\tilde{\beta}_{-}v^2)k^2=0.
\end{equation}
The dispersion relation can be easily found and reads
\begin{equation}
\omega=\frac{-\tilde{\beta}_{-}\tilde{\beta}_{+}(\vec{v}\cdot\vec{k})\pm c_{s}|k|\sqrt{\tilde{\beta}_{-}\tilde{\beta}_{+}(\tilde{\beta}_{+}+c^2_{s}-\tilde{\beta}_{-}v^2})}{\tilde{\beta}_{+}(\tilde{\beta}_{+}+c^2_{s})}.
\end{equation}
This relation is different from the result obtained in \cite{MV} and \cite{Xian}.
In the limit $c^2_{s}\ll1$ and $v^2\ll1$ the dispersion relation is simply given by
\begin{equation}
\omega\approx\frac{-\tilde{\beta}_{-}(\vec{v}\cdot\vec{k})}{\tilde{\beta}_{+}}\pm\frac{c_{s}\sqrt{\tilde{\beta}_{-}}}{\tilde{\beta}_{+}}|k|
=\frac{-(1-\beta)(\vec{v}\cdot\vec{k})}{1+\beta}\pm\frac{c_{s}\sqrt{1-\beta}}{1+\beta}|k|.
\end{equation}
Notice that we obtain a correction to the dispersion relation and for $\beta=0$, we recover the result obtained in \cite{Xian}.   
On the other hand, in the regime $c_s^2=1\gg v^2$, for $0\leq\beta<1$, one finds
\begin{equation}
\label{group}
\omega=\pm \sqrt{\frac{1-\beta}{1+\beta}}|k|.
\end{equation}
Here the group speed $v_g=\frac{d\omega}{d|k|}=\pm \sqrt{\frac{1-\beta}{1+\beta}}$ is such that $|v_g|<c_s$. The Lorentz invariance usually
governing
the sound waves (the fluid fluctuations) here is broken because the Lorentz boost invariance now is lost.
This does not produces birefringence though, since in any direction the group speed has the same value $|v_g|$. One alternative to achieve the birefringence phenomenon would be considering now $\beta=0$ and $\alpha\neq0$ in the tensor $k_{\mu\nu}$ as in Ref.~\cite{Bazeia:2005tb}. 
In the next subsection, we shall address this issue.

\subsection{The case $\beta=0$ and $\alpha\neq 0$}
\noindent
In terms of momenta and frequency the equation (\ref{eqkga}) becomes 
\begin{eqnarray}
[c^2_{s}+(1-\alpha v)^2]\omega^2+\left[2(1-\alpha v)(\vec{v}\cdot\vec{k})\right]\omega-[(1+\alpha)c^2_{s}-v^2+\alpha^2(1-v)^2]k^2=0.
\end{eqnarray}
The dispersion relation can be easily found and reads
\begin{equation}
\omega=\frac{-\left[2(1-\alpha v)(\vec{v}\cdot\vec{k})\right] \pm \sqrt{\Delta}}{2[c^2_{s}+(1-\alpha v)^2]},
\end{equation} 
where
\begin{eqnarray}
\Delta&=&\left[2(1-\alpha v)(\vec{v}\cdot\vec{k})\right]^2
+4[c^2_{s}+(1-\alpha v)^2][(1+\alpha)c^2_{s}-v^2+\alpha^2(1-v)^2]k^2,
\\
&=&4c^2_{s}k^2\left[1+c^2_{s}-v^2+\alpha(1+ c^2_{s})+\alpha^2(1-v)^2+\alpha v(\alpha v-2)(1+\alpha)\right.
\nonumber\\
&+&\left.\frac{\alpha^2(1-v)^2(1-\alpha v)}{c^2_{s}}\right].
\end{eqnarray}
In the limit $c_s^2=1\gg v^2$ and $0\leq\alpha\leq1$ we find the simple dispersion relation
\begin{eqnarray}
\omega=-\frac{1}{2} \cos{\theta}|k|v\pm{\frac {\sqrt {2} \left( 2+2\alpha+2{\alpha}^{2}-\,{
\alpha}^{2}v+\alpha v+{\alpha}^{3}v \right) |k|}{4\sqrt{1+\alpha+{
\alpha}^{2}}}}.
\end{eqnarray}
We can easily check that the group speed $v_g^{\pm}$ (for $v\neq0$) has different intensities at different directions. This 
clearly breaks the Lorentz invariance and produces the effect of birefringence mentioned in the previous section. 

\section{Lorentz violating canonical acoustic black hole}
\label{IV}
In this section, we shall address the issue of Hawking temperature in the regime of low velocities 
for the previous cases with further details. Now we consider an incompressible fluid with spherical symmetry. In this case the density $\rho$ is a position independent quantity and the continuity equation implies that $v\sim \frac{1}{r^2}$. The sound speed is also a constant.

\subsection{The case $\beta\neq0$ and $\alpha=0$}
\noindent
The Lorentz violating acoustic metric 
in the limit $c^2_{s}\ll1$ and $v^2\ll1$ can be written as a Schwarzschild metric type, up to an irrelevant position-independent factor, as follows,
\begin{eqnarray}
ds^2&=&-f(v_{r})d\tau^2+\frac{c^2_{s}}{\sqrt{\tilde{\beta}_{-}\tilde{\beta}_{+}}f(v_{r})}dr^2
+\sqrt{\frac{\tilde{\beta_{+}}}{\tilde{\beta}_{-}}}r^2(d\theta^2+\sin^2\theta d\phi^2),
\end{eqnarray}
where
\begin{equation}
\label{hz1}
f(v_{r})=\sqrt{\frac{\tilde{\beta_{-}}}{\tilde{\beta}_{+}}}\left[\frac{c^2_{s}-\tilde{\beta}_{-}v^2_{r}}{\tilde{\beta}_{+}}\right]\rightarrow f(r)=\sqrt{\frac{\tilde{\beta_{-}}}{\tilde{\beta}_{+}}}\left[\frac{c^2_{s}}{\tilde{\beta}_{+}}(1-\tilde{\beta}_{-}\frac{r^4_{h}}{r^4})\right].
\end{equation}
We also have defined
$v_{r}=c_{s}\frac{r^2_{h}}{r^2}$,
where $r_{h}$ is the event horizon, the radius at
which the flow speed exceeds the sound speed in the fluid. 

The Hawking temperature is given by
\begin{equation}
T_{H}=\frac{f^{\prime}(r_{h})}{4\pi}=c^2_{s}\tilde{\beta}^{3/2}_{-}\tilde{\beta}^{-3/2}_{+}(\pi r_{h})^{-1}
=\frac{c^2_{s}(1-\beta)^{3/2}}{(1+\beta)^{3/2}\pi r_{h}}.
\end{equation}
For $\beta=0$ the usual result is obtained. Note that $\beta$ plays the role of changing the intensity of the Hawking temperature. It is maximal as $\beta\to-1$ and minimal as $\beta\to1$. 
Thus, the parameter $\beta$ affects directly the temperature of the acoustic black hole. 

Furthermore, it is worth noting that the Hawking temperature can be written in terms of the group speed obtained from (\ref{group}) in the form
\begin{equation}
T_{H}\to T_{H}^{eff}=v_g^3 T_H, \qquad {\rm where} \qquad v_g^3=\frac{c^2_{s}(1-\beta)^{3/2}}{(1+\beta)^{3/2}}.
\end{equation}
One can make an analogy with Lorentz-violating gravitational black holes \cite{syb,adam} as in the following. This way of the relating the 
{\it effective Hawking temperature} $T_{H}^{eff}$ means that the Hawking temperature is {\it not} universal for all species of particles. It
depends on the maximal attainable velocity of this species. 
In the context of gravitational black holes this has been previously studied and appointed
as a sign of possibly violation of the second law of the thermodynamics \cite{syb}.

Another solution for the fluid flow is considering that at some value of $r=r_{+}$ (sonic horizon) we have the background fluid smoothly exceeding the sound speed, so the velocity $v(r)$ can be expanded as
$v_{r}=-\tilde{\beta}_{-}^{-1/2}c_{s}+a(r-r_{+})+O(r-r_{+})^2$,
which yields, up to the first order in $r$, 
$c^2_{s}-\tilde{\beta}_{-}v^2_{r}\approx 2\tilde{\beta}_{-}^{1/2}ac_{s}(r-r_{+})$.
In this limit our metric becomes
\begin{eqnarray}
ds^2&=&-\frac{2\tilde{\beta}_{-}ac_{s}(r-r_{+})}{\tilde{\beta}^{3/2}_{+}}d\tau^2
+\frac{\tilde{\beta}_{+}c_{s}}
{2a\tilde{\beta}^{3/2}_{-}(r-r_{+})}dr^2
+\sqrt{\frac{\tilde{\beta_{+}}}{\tilde{\beta}_{-}}}r^2(d\theta^2+\sin^2\theta d\phi^2),
\end{eqnarray}
where $a$ is a parameter associated with the velocity of the fluid defined as $(\nabla\cdot\vec {v})|_{r=r_{+}}$.
The Hawking temperature for the acoustic black hole is now given by
\begin{equation} 
T_{H}=\frac{\tilde{\beta}_{-}ac_{s}}{2\pi\tilde{\beta}^{3/2}_{+}}=\frac{(1-\beta)ac_{s}}{2\pi(1+\beta)^{3/2}}.
\end{equation}
This result coincides, for $\beta=0$, with the result of the Hawking temperature obtained in \cite{Kim} and the thermal
emission is proportional to the control parameter $a$. The behavior of the Hawking temperature with respect to $\beta$ discussed
in the previous case also applies here.

\subsection{The case $\beta=0$ and $\alpha\neq 0$}

In the limit $c^2_{s}\ll1$ and $v^2\ll1$, we have
\begin{eqnarray}
&&f(v_{r})=\frac{[(1+\alpha)c^2_{s}-v^2+\alpha^2(1-v)^2]}{\sqrt{\tilde{\alpha}(1-2\alpha v_{r})+\alpha^2[(1-2v_{r})+c^{-2}_{s}(1-2v_{r}\tilde{\alpha})]}} 
\nonumber\\
&&
\rightarrow f(r)=\frac{c^2_{s}\left[\tilde{\alpha}-\frac{r^4_{h}}{r^4}\right]+\alpha^2\left(1-c_{s}\frac{r^2_{h}}{r^2}\right)^2}
{\sqrt{\tilde{\alpha}\left(1-2\alpha c_{s}\frac{r^2_{h}}{r^2}\right)
+\alpha^2\left[\left(1-2c_{s}\frac{r^2_{h}}{r^2}\right)+c^{-2}_{s}\left(1-2c_{s}\frac{r^2_{h}}{r^2}\tilde{\alpha}\right)\right]}},
\end{eqnarray}
where $\tilde{\alpha}=1+\alpha$. For $\alpha$ sufficiently small we have up to first order
\begin{equation}
f(r)\simeq c_s^2\,\tilde{\alpha}\left(1-\frac{1}{\tilde{\alpha}}\frac{r_h^4}{r^4}\right)
+c_s^3\,\left(\frac{r_h^2}{r^2}-\frac{1}{2c_{s}}\right)\left(1-\frac{r_h^4}{r^4}\right)\alpha.
\end{equation}
In the present case there is a richer structure such as {\it charged} and {\it rotating} black holes. 
The Hawking temperature is
\begin{equation}
T_{H}=\frac{[c^2_{s}+\alpha^2 c_{s}(1-c_{s})]}
{\pi r_{h}\sqrt{\Lambda}}
-\frac{\alpha^2 c_{s}[c^2_{s}+\alpha(1-c_{s})^2]}{2\pi r_{h}(\Lambda)^{3/2}},
\end{equation}
where $\Lambda=\tilde{\alpha}\left(1-2\alpha c_{s}\right)
+\alpha^2\left[\left(1-2c_{s}\right)+c^{-2}_{s}\left(1-2c_{s}\tilde{\alpha}\right)\right]$. For $\alpha=0$ one recovers the usual result. One can make a discussion with several regimes with such temperature.
For instance, for $(i)$ $\alpha\ll1$ and $c_s^2=1$ up to first order in $\alpha$ we have:
$T_{H}=\left(1+\frac12\alpha\right)/(\pi r_h)\to T_H^{eff}\sim v_g T_H$,
and for $(ii)$ $\alpha\to\frac{\sqrt{3}-1}{2}$ and $c_s^2=1$ the Hawking temperature diverges as
\begin{equation}
T\to \frac{1}{r_{h_{eff}}}\sim\frac{1}{(-\alpha+\frac{\sqrt{3}-1}{2})^{3/2}r_h}.
\end{equation}
The {\it effective horizon} $r_{h_{eff}}$ goes to zero whereas the {\it effective curvature} diverges. Notice that in the latter case 
small variations of $\alpha$ give effects larger than in the former case.
In the gravitational context, it has been suggested that the effects of the Lorentz symmetry violation may be larger
in regions of large curvature and torsion such as black holes \cite{Kostelecky:2003fs}.

Both of the cases studied in this section comprise spherically symmetric black hole solutions for slow speeds $c_s\ll1$ and 
$v\ll1$. However, exploring the original solutions (\ref{invs_g}) and (\ref{met}) as spherically symmetric solutions with 
$v_z=0$, $v_r\neq0$ and $v_\phi\neq0$ one
can show that they can be written in a Kerr-like form
\begin{eqnarray}
\label{kerr}
ds^2&=&\frac{b\rho_0}{2c_s}\left[-N^2 d\tau^2+M^2 dr^2+Q^2 r^2d\varphi^2+Z^2dz^2
+\frac{(\Theta v_{\phi}d\tau -\Omega rd\varphi)^2}{\sqrt{f}}\right].
\end{eqnarray}
Thus, let us write the acoustic black hole metric of the previous case in the Kerr-like form. Firstly, consider the case 
$\beta\neq0$ and $\alpha=0$, where we have the Kerr-like components as 
\begin{equation}
\label{kerr2}
N^2(v_r,v_\phi)=\frac{V_N}{\sqrt{f}},
\quad
M^2(v_r,v_\phi)=\frac{c^2_{s}}{N^2(v_r,v_\phi)},
\quad
Q^2(v_r,v_\phi)=\frac{\left(\frac{c_{s}^2}{\tilde{\beta}_{-}}-\frac{\tilde{\beta}_{-}}{\tilde{\beta}_{+}}v^2_{r}\right)}
{\sqrt{f}},
\quad
Z^2(v_r,v_\phi)=\frac{f_{\beta}}{\sqrt{f}}
\end{equation}
\begin{equation}
f_{\beta}=\frac{\tilde{\beta}_{+}}{\tilde{\beta}_{-}}+\frac{c_{s}^2}{\tilde{\beta}_{-}}-\frac{\tilde{\beta}_{-}}{\tilde{\beta}_{+}}v^2,
\quad f=\frac{1}{\tilde{\beta}_{-}}\left(1+\frac{c_{s}^2}{\tilde{\beta}_{+}}
-\frac{\tilde{\beta}_{-}}{\tilde{\beta}_{+}}v^2_{r}\right),\quad\Theta=\sqrt{\frac{\tilde{\beta}_{-}}{\tilde{\beta}_{+}}},
\quad\Omega=\sqrt{\frac{\tilde{\beta}_{+}}{\tilde{\beta}_{-}}},
\end{equation}
where $V_N=\left(\frac{c_{s}^2}{\tilde{\beta}_{+}}-\frac{\tilde{\beta}_{-}}{\tilde{\beta}_{+}}v^2_{r}\right)$ and the coordinate transformations we have been used are
\begin{eqnarray}
d\tau=dt+\frac{v_{r}dr}{\left(\frac{c_{s}^2}{\tilde{\beta}_{+}}-\frac{\tilde{\beta}_{-}}{\tilde{\beta}_{+}}v^2_{r}\right)},\quad\quad
d\varphi=d\phi+\frac{\tilde{\beta}_{-}v_{r}v_{\phi}dr}{\tilde{\beta}_{+}r\left(\frac{c_{s}^2}{\tilde{\beta}_{-}}-\frac{\tilde{\beta}_{-}}
{\tilde{\beta}_{+}}v^2_{r}\right)}.
\end{eqnarray}
Now we consider the Kerr-like form for second case, i.e. $\beta=0$ and $\alpha\neq 0$, to find the following components
\begin{equation}
N^2(v_r,v_\phi)=\frac{(1+\alpha)c^2_{s}-v^2_{r}+\alpha^2(1-v)^2}{\sqrt{f}},
\quad
M^2(v_r,v_\phi)=\frac{c^2_{s}}{N^2(v_r,v_\phi)},
\quad
Q^2(v_r,v_\phi)=\frac{c^2_{s}-v^2_{r}}{\sqrt{f}},
\end{equation}
\begin{equation}
Z^2(v_r,v_\phi)=\frac{f_{\alpha}}{\sqrt{f}},
\quad
f=(1+\alpha)[(1-\alpha v)^2+c^2_{s}]-v^2+\alpha^{2}(1-v)^2\left[1+(1-\alpha v)^2c^{-2}_{s}\right],
\end{equation}
\begin{equation}
f_{\alpha}=(1-\alpha v)^2+c^2_{s}-v^2_{r},\quad \Theta=1,\quad\Omega=(1-\alpha v).
\end{equation}
The coordinate transformations we have been used in this case reads
\begin{eqnarray}
d\tau=dt+\frac{(1-\alpha v)v_{r}dr}{\left[(1+\alpha)c^2_{s}-v^2_{r}+\alpha^2(1-v)^2\right]},\quad
d\varphi=d\phi+\frac{v_{r}v_{\phi}dr}{r\left(c_{s}^2-v^2_{r}\right)}.
\end{eqnarray}
In the limit of $\alpha,\beta=0$ one recovers the relativistic result of \cite{Xian}.
Note that the Lorentz-violating parameters $\alpha,\beta$ affect the components of the Kerr-like metric and as a consequence the horizons also 
change. Thus that in any case the Lorentz violating term acting directing on the fluid also affects the acoustic black hole temperature.

The fact that we have found a Kerr-like metric in our system is not surprising because as one knows the Abelian Higgs Model 
exhibits vortex solutions. As a consequence we are indeed dealing with the ``draining bathtub'' fluid flow \cite{Volovik,MV}. The Lorentz
violating term affects the ergosphere and acoustic event horizon as follows. Let us illustrate this by using the case $\beta\neq0,\alpha=0$. We can read off from Eqs.~({\ref{kerr}})-(\ref{kerr2}) 
\begin{equation}
g_{\tau\tau}=
\frac{\frac{c_{s}^2}{\tilde{\beta}_{+}}-\frac{\tilde{\beta}_{-}}{\tilde{\beta}_{+}}(v^2_{r}+v^2_{\phi})}{\sqrt{f}}, \qquad g_{rr}=
c_s^2\left(\frac{\frac{c_{s}^2}{\tilde{\beta}_{+}}-\frac{\tilde{\beta}_{-}}{\tilde{\beta}_{+}}v^2_{r}}{\sqrt{f}}\right)^{-1}.
\end{equation}
Let us now use the velocity components $v_r=\frac{A}{r}$ and $v_\phi=\frac{B}{r}$ that follow from equation of continuity and 
conservation of angular momentum, respectively. The radius of the ergosphere is given by $g_{00}(r_e)=0$, whereas the horizon
is given by the coordinate singularity $g_{rr}(r_h)=0$, that is
\begin{equation}
r_e=\frac{\tilde{\beta}_-^{1/2}(A^2+B^2)^{1/2}}{c_s}, \qquad r_h=\frac{\tilde{\beta}_-^{1/2}|A|}{c_s}.
\end{equation}
Many interesting studies can be followed from this point. One of them is the `superresonance' \cite{Basak:2002aw} which is an analog of the 
superradiance phenomenon in gravitational black holes, but a detailed study on this subject is out of the scope of this paper. We 
shall consider this study in a forthcoming publication. However we can anticipate some few issues as follows. 
The scattering of a particle wave function by the Kerr-like acoustic black roles gives arise to amplitude increasing. This is well-known as the
Penrose process. Here the amplitude of a scattered wave function is determined by the reflectance $|{\cal R}|$ given as 
\begin{equation}
\label{omega_H}
1-|{\cal R}|^2=\left(\frac{\omega-m\tilde{\Omega}_H}{\omega}\right)|{\cal T}|^2,\qquad \tilde{\Omega}_H=\frac{\Omega_H}{\tilde{\beta}_-^{1/2}}.
\end{equation}
For frequencies in the interval $0<\omega<m\tilde{\Omega}_H$ the reflectance is always larger than unit, which implies in the superresonance phenomenon. Here $m$ is the azimuthal mode number and $\Omega_H=Bc_s/A^2$ is the angular velocity of the usual Kerr-like acoustic black hole.
Notice from Eq.~(\ref{omega_H}) that the modified angular velocity $\tilde{\Omega}_H$ depends on the Lorentz violating parameter $\tilde{\beta}_-=1-\beta$. This means that for $\beta\to1$ a larger spectrum of particles wave function can be scattered with increased amplitude.
Alternatively, the acoustic Kerr-like black hole possesses a wider rate of loss of mass (energy).

\section{(2+1) Dimensional Model}
\label{V}
In this section, let us make a few comments on a (2+1) dimensional model. We apply our previous analysis to the (2+1)-dimensional spacetime. The Maxwell's term is now replaced to the Chern-Simons term, but the Lorentz-violating term early considered is still applied in the present discussion.
\subsection{The case $\beta=0$ and $\alpha\neq 0$}
\noindent
To describe the theory in (2+1) dimensions Lorentz-violating we consider the Lagrangian given by
\begin{eqnarray}
{\cal L}&=&\frac{\kappa}{2} \epsilon^{\mu\nu\lambda}(A_{\mu}\partial_{\nu} A_{\lambda})+ |D_{\mu}\phi|^2+ m^2|\phi|^2-b|\phi|^4 
+ k^{\mu\nu}D_{\mu}\phi^{\ast}D_{\nu}\phi,
\end{eqnarray}
and using the decomposition $\phi=\sqrt{\rho(x,t)}\exp{(iS(x,t))}$, we obtain 
\begin{eqnarray}
{\cal L}&=&\frac{\kappa}{2} \epsilon^{\mu\nu\lambda}(A_{\mu}\partial_{\nu} A_{\lambda})+\rho\partial_{\mu}S\partial^{\mu}S
-2e\rho A_{\mu}\partial^{\mu}S + e^2\rho A_{\mu}A^{\mu} + m^2\rho-b\rho^2
\nonumber\\
&+&k^{\mu\nu}\rho(\partial_{\mu}S\partial_{\nu}S-2eA_{\mu}\partial_{\nu}S+ e^2 A_{\mu}A_{\nu})
+\frac{\rho}{\sqrt{\rho}}(\partial_{\mu}\partial^{\mu}+k^{\mu\nu}\partial_{\mu}\partial_{\nu})\sqrt{\rho}.
\end{eqnarray}
In this case, the (2+1)-dimensional metric has the same components of the metric (\ref{met}) in (3+1) dimensions, up to an overall constant factor, and reads 
\begin{equation}
g_{\mu\nu}\equiv\frac{b\rho^2_{0}}{2c^2_{s}\sqrt{f}}
\left[\begin{array}{clcl}
g_{tt} &\vdots &g_{tj}\\
\cdots&\cdot&\cdots\\
g_{it} &\vdots &g_{ij}
\end{array}\right],
\end{equation}
where $g_{tt},g_{ti},g_{it},g_{ij}$ is given by equations (\ref{gtt})-(\ref{gij}).

The equations of motion for $A_{0}$ and $A_{i}$ are:
\begin{equation}
\label{lg}
B=\epsilon^{ij}\partial_{i}A_{j}=\partial_{1}A_{2}-\partial_{2}A_{1}=\frac{2e\rho}{\kappa}[\partial_{0}S-eA_{0}+k^{0i}(\partial_{i}S-eA_{i})],
\end{equation}
\begin{equation}
\partial_{0}A_{j}-\partial_{j}A_{0}=2e\rho\epsilon_{ji}[(\partial^{i}S-eA^{i})- k^{0i}(-\partial_{0}S+eA_{0})+k^{il}(\partial_{l}S-eA_{l})],
\end{equation}
where the equation (\ref{lg}) is the Chern-Simons Gauss law. 
Now we can completely exclude from consideration the vector potentials $A_{\mu}$ in favor of the local velocity field
\begin{eqnarray}
\label{rot}
\nabla\times\vec{v}&=&\frac{2e^2\rho}{\kappa}[1-\alpha v],
\\
\partial_{0}A_{j}-\partial_{j}A_{0}&=&2e\rho\epsilon_{ji}w_{0}[v^{i}- k^{0i}+k^{il}v_{l}],
\end{eqnarray}
where $v=v_{0}/w_{0}$. The equation (\ref{rot}) has the simple meaning of a rotational fluid.

\subsection{The case $\beta\neq 0$ and $\alpha=0$}
\noindent
In this case, the equations of motion for $A_{0}$ and $A_{i}$ are
\begin{equation}
B=\epsilon^{ij}\partial_{i}A_{j}=\partial_{1}A_{2}-\partial_{2}A_{1}
=\frac{2e\rho}{\kappa}\tilde{\beta}_{+}(\partial_{0}S-eA_{0}),
\end{equation}
\begin{equation}
\partial_{0}A_{j}-\partial_{j}A_{0}=2e\rho\epsilon_{ji}\tilde{\beta}_{-}(\partial^{i}S-eA^{i}),
\end{equation}
and in terms of the local velocity field
\begin{eqnarray}
\nabla\times\vec{v}&=&\frac{2e^2\rho}{\kappa}(1+\beta),
\\
\partial_{0}A_{j}-\partial_{j}A_{0}&=&2e\rho\epsilon_{ji}(1-\beta)v^{i}_{0}.
\end{eqnarray}
Again, in the present case the (2+1)-dimensional metric has the same components of the metric (\ref{invs_g}) in (3+1) dimensions, up to an overall constant factor, and is given by
\begin{equation}
g_{\mu\nu}\equiv\frac{b\rho^{2}_{0}\sqrt{\tilde{\beta}_{-}}}{2c^{2}_{s}\sqrt{1+\frac{c_{s}^2}{\tilde{\beta}_{+}}
-\frac{\tilde{\beta}_{-}}{\tilde{\beta}_{+}}v^2}}
\left[\begin{array}{clcl}
-\left(\frac{c_{s}^2}{\tilde{\beta}_{+}}-\frac{\tilde{\beta}_{-}}{\tilde{\beta}_{+}}v^2\right)&\vdots &\quad\quad\quad\quad-v^{j}\\
\cdots\cdots\cdots\cdots\cdots&\cdot&\quad\quad\cdots\cdots\cdots\cdots\cdots\cdots\\
-v^{i}&\vdots & \left(\frac{\tilde{\beta}_{+}}{\tilde{\beta}_{-}}+\frac{c_{s}^2}{\tilde{\beta}_{-}}
-\frac{\tilde{\beta}_{-}}{\tilde{\beta}_{+}}v^2\right)\delta^{ij}+\frac{\tilde{\beta}_{-}}{\tilde{\beta}_{+}}v^{i}v^{j}
\end{array}\right].
\end{equation}
This example shows how analogous but different is the acoustic metric compared to gravitational metrics. In the gravitational realm the dimension of the
spacetime radically affects the gravitational solutions such as black holes.

\section{Conclusions}
\label{conclu}

In this paper we have considered the extended Abelian Higgs model with a Lorentz-violating term. The consequences are that the acoustic Hawking temperature is changed such that it depends on the group speed which means that, analogously to the gravitational case \cite{syb,adam}, the Hawking temperature is {\it not} universal for all species of particles. It
depends on the maximal attainable velocity of this species. In the context of gravitational black holes this has been previously studied and appointed
as a sign of possibly violation of the second law of the thermodynamics.  Furthermore, the acoustic black hole metric in our model can be identified with 
an acoustic Kerr-like black hole. The Lorentz violating term affects the rate of loss of mass (energy) of the black hole. We also have shown that for suitable values of
the Lorentz violating parameter a wider spectrum of particle wave function can be scattered with increased amplitude by the acoustic black hole. This
increases the superressonance phenomenon previously studied in \cite{Basak:2002aw}. On the other hand, the Abelian Higgs model is good to describe high energy physics
and extended Abelian Higgs model can also describe Lorentz symmetry violation in particle physics in high energy. Thus our results show that in addition to the expected gravitational mini black holes formed in high energy experiments one can also expect the formation of acoustic black holes together.

\acknowledgments

We would like to thank L. Barosi for discussions and CNPq, CAPES, PNPD/PROCAD -
CAPES for partial financial support.

\end{document}